\documentclass[a4paper,citeautoscript,amsmath,amssymb,reprint,11pt]{article}
\usepackage[top=2.5cm,bottom =3cm,left = 2cm, right = 2cm]{geometry}

\usepackage{graphicx}
\usepackage{wrapfig}
\usepackage{subcaption}
\usepackage{epstopdf}
\usepackage{float}
\usepackage{siunitx}
\usepackage{amsmath,amsfonts,amsthm,bm}

\newcommand{\etal}{\textit{et al}.\ }
\DeclareSIUnit\dBm{dBm}
\DeclareSIUnit\dB{dB}
\DeclareSIUnit\mbar{mbar}

\begin{document}
\sloppy
\begin{center}

\textbf{{\large Experimental verification of electrostatic boundary conditions in gate-patterned quantum devices}}\\

Hou. H$^{1*}$, Y. Chung$^{1(}$\footnote{Present address:
Centre for Quantum Computation and Communication Technology, School of Physics, The University of New South Wales, Sydney, Australia}$^)$, G. Rughoobur$^{2(}$\footnote{Present address:
Microsystems Technology Laboratories, Massachusetts Institute of Technology, Massachusetts, USA}$^)$, T. K. Hsiao$^1$, A. Nasir$^{1,3}$, A. J. Flewitt$^2$, J. P. Griffiths$^1$, I. Farrer$^{1(}$\footnote{Present address:
Electronic and Electrical Engineering Department, The University of Sheffield, Sheffield, UK}$^)$, D. A. Ritchie$^1$, C. J. B. Ford$^1$\\

$ˆ1$ Cavendish Laboratory, University of Cambridge, Cambridge, CB3 0HE, UK\\
$  ˆ2$ Department of Engineering, University of Cambridge,Cambridge, CB3 0FA, UK\\
$ˆ3$ National Physical Laboratory, Hampton Road, Middlesex, TW11 0LW, UK\\

   $*$ hh405@cam.ac.uk \\

\begin{abstract}
In a model of a gate-patterned quantum device it is important to choose the correct electrostatic boundary conditions (BCs) in order to match experiment. In this study, we model gated-patterned devices in doped and undoped GaAs heterostructures for a variety of BCs. The best match is obtained for an unconstrained surface between the gates, with a dielectric region above it and a frozen layer of surface charge, together with a very deep back boundary. Experimentally, we find a $\sim\SI{0.2}{V}$ offset in pinch-off characteristics of one-dimensional channels in a doped heterostructure before and after etching off a ZnO overlayer, as predicted by the model. Also, we observe a clear quantised current driven by a surface acoustic wave through a lateral induced $n$-$i$-$n$ junction in an undoped heterostructure. In the model, the ability to pump electrons in this type of device is highly sensitive to the back BC. Using the improved boundary conditions, it is straightforward to model quantum devices quite accurately using standard software.
\end{abstract}
\end{center}
\twocolumn
\section{Introduction}
Gate-patterned devices using a two-dimensional electron gas (2DEG)
allow investigation of a variety of effects such as ballistic electron transport,\cite{wharam1988one,1D,thomas1996possible} Coulomb blockade \cite{livermore1996coulomb,ono2002current} and spin read-out \cite{elzerman2004single,hanson2005single}, and they are being developed for their use in quantum computation. To understand the shape of the potential in such devices, and to optimise designs, it is essential to calculate the electrostatic potential distribution with specific patterned gates and various biases. However, the most realistic surface and back boundary conditions (BCs) are still controversial despite much work over the years.\cite{ laux1988quasi,kumar1990electron,snider1990electron,chen1994coupled,chen1995design,davies1995modeling,sun1995electrostatic,lier1993self,fiori2002experimental}
In solving Poisson's equation $\nabla\cdot(\varepsilon\varepsilon_0\nabla \phi)=-\rho$, one must specify on the boundary either the electrostatic potential $\phi$ (Dirichlet BCs) or its normal derivative $\partial \phi/\partial n$ (Neumann BCs), or a mixture of the two. For GaAs, the high density of surface-charge states pins the Fermi level at the surface near the middle of the band gap, $\sim\SI{0.75}{eV}$ below the conduction band minimum. At high temperatures the charge is mobile and there is no difference between a gated surface and an exposed surface. At cryogenic temperatures, the surface charge does not vary as gate biases are changed i.e.\ it is `frozen', because the temperature is too low (below \SI{100}{K}) for charge to move out of traps in the donor layer or at the surface. If this were not the case, then split-gate devices would exhibit hysteresis or pinch off gradually over time, as the charge hopped between surface states. This is not observed for 2DEGs, although for hole gases, charge may be able to move between acceptors because it is less tightly bound.\cite{Daneshvar1997-1936} 

Thus, if surface gates are varied while the device is cold, the exposed surface will no longer be an equipotential, though this is still a popular approximation as it simplifies the calculation.\cite{snider1990electron,guo2009threeold} Chen \etal considered similar surface BCs previously,\cite{chen1994coupled,chen1995design} and devised a sophisticated scheme to include the `air' above the surface. They showed that Neumann BCs on the surface matched the full calculation with air well, and gave much better results than using Dirichlet BCs. However, they did not consider the case of a surface dielectric instead of air. Here, we find that this layer causes a significant shift in the pinch-off voltage. We also apply the idea of frozen charge below the 2DEG, which is ignored in the above studies. There are interface states at the `dirty' regrowth interface between the substrate and the heterostructure grown by molecular-beam epitaxy (MBE), which can be calculated from the built-in electrical field arising from intentional and unintentional dopants. These charge states also freeze out, so this interface is no longer at a constant potential. Therefore, instead of using Dirichlet BCs at the regrowth interface, as is often done, we take a fictitious boundary significantly below the regrowth interface.

In this work we calculate three-dimensional electrostatic potentials using a standard commercial partial differential equation (PDE) solver package, \textit{Nextnano}\cite{nextnano}. We compare simulations with a range of experimental verifications, doped and undoped GaAs-based heterostructures, and patterned-gate structures with and without a surface dielectric layer. We find that the models match experiments significantly more closely, and have much greater predictive power in device architectures, if one chooses the BCs carefully.   

\section{Boundary conditions in doped gated devices}
\begin{figure}
\begin{center}
\captionsetup{justification=raggedright}
\includegraphics[scale=0.6]{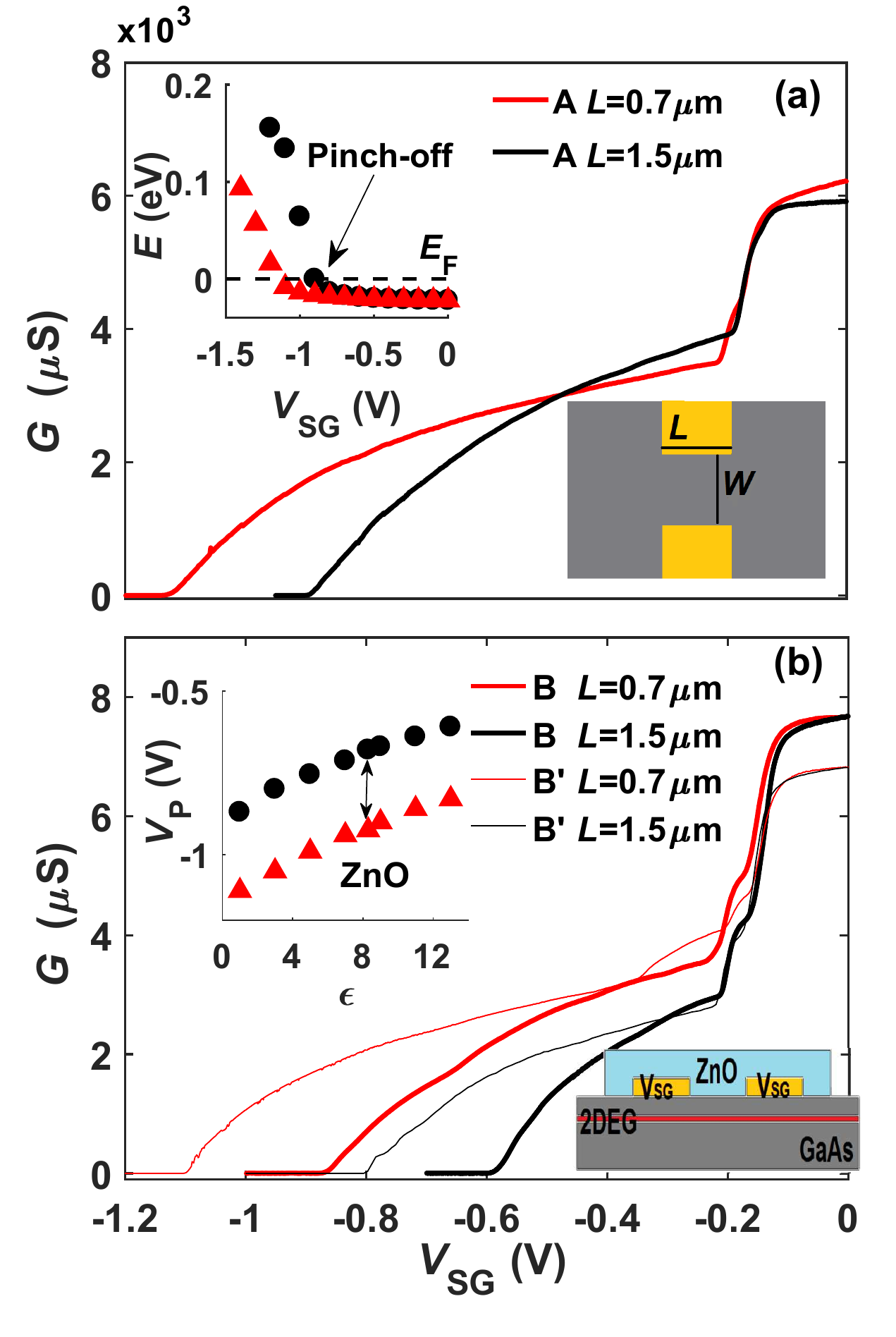}
\end{center}
\caption{\label{modelpinch}(a) Schematic diagram of a 1D channel and pinch-off characteristics in chip A. Inset: the conduction-band minimum of the centre of a 1D constriction of dimensions $L=\SI{0.7}{\mu\meter}$ (red triangles) or $L=\SI{1.5}{\mu\meter}$ (black circles) as a function of $V_{\rm SG}$.  The Fermi level $E_{\rm F}$ is taken to be 0. (b) Schematic diagram of chip B with a ZnO overlayer and the pinch-off characteristics with (thick line) and without (thin line) ZnO overlayer. Inset: calculated pinch-off voltage of a 1D channel with $L=\SI{1.5}{\mu\meter}$ (black circles) or $L=\SI{0.7}{\mu\meter}$ (red triangles) as a function of dielectric constant.}
\end{figure}

The first structure we model is a pair of split gates defining a narrow one-dimensional (1D) channel in the 2DEG of a Si-doped GaAs/AlGaAs heterostructure when a negative bias $V_{\rm SG}$ is applied to the gates. For a 2DEG density of \SI{1.5e11}{cm^{-2}}, we calculate the surface (back) charge density \SI{-2.18e12}{cm^{-2}} (\SI{5.45e10}{cm^{-2}})  at about \SI{100}{K}, and include it at low temperatures as fixed $\delta$-doping layers at the front surface and regrowth interfaces. Then the electrostatic potential distribution across the 1D channel is mapped as a function of $V_{\rm SG}$. Instead of the direct BCs on the exposed surface, a vacuum layer (a material with large bandgap \SI{15}{eV} and dielectric constant $\varepsilon=1$) is introduced with Neumann BC $\partial \phi/\partial z=0$ above the vacuum. As discussed above, the back BCs should be applied deep below the regrowth interface. However, for these doped structures we find no significant effect of moving the boundary from this interface. This is because, for there to be charge in the 2DEG, there is a large built-in field above the interface even at high temperature, and this is screened by the charge at the regrowth interface, which then becomes frozen. Changing a surface gate voltage then causes a relatively minor shift in the bands, and taking the lower boundary any distance below the regrowth interface gives similar results. As $V_{\rm SG}$ becomes negative, the conduction-band minimum at the centre of the 1D channel starts to rise. When it is above the Fermi level, the electrons become fully depleted, pinching off the channel at $V_{\rm SG}=V_{\rm P}$, as shown in the inset to Fig.~\ref{modelpinch}(a) for two different channel lengths $L$ ($0.7$ and $\SI{1.5}{\mu\meter}$) but the same width $W$ ($\SI{0.7}{\mu\meter}$). We find that there is no significant difference (within 2\%) in both $V_{\rm P}$ and the confining potential in the channel between the case where the Neumann BC is applied on the exposed surface, and the case with a vacuum layer. This is because the large difference in dielectric constant causes electric fields just inside the surface to be nearly parallel to it, as for the Neumann BC. If, instead, the vacuum region is replaced by a different dielectric layer with $\varepsilon\gg 1$, the result is different. The inset to Fig.~\ref{modelpinch}(b) demonstrates the increasing effect of the dielectric layer as $\varepsilon$ is varied from 1 (vacuum) to 13 (GaAs). For example, for ZnO ($\varepsilon=8.3$), $V_{\rm P}$ shifts by about \SI{0.2}{V}, which should be observable in experiments.

\begin{table*}
\centering
	\begin{tabular}{||c|c|c|c||} 
		\hline
		Dielectric & $L$ ($\mu$m) & $V_{\rm P}$: experiment (V) & $V_{\rm P}$: model (V) \\ [0.5ex] 
		\hline
		Vacuum & 0.7 & $-1.15\pm0.05$ & $-1.16\pm0.02$\\
		Vacuum & 1.5 & $-0.89\pm0.05$ & $-0.9\pm0.02$\\
		ZnO & 0.7 & $-0.87\pm0.03$ & $-0.92\pm0.03$\\
		ZnO & 1.5 & $-0.60\pm0.03$ & $-0.68\pm0.03$\\
		\hline
	\end{tabular}
	\caption{\label{tab:table1}Comparison of experimental and modelled pinch-off voltages.}
\end{table*}
\section{1D channel pinch-off characterisation}
For a comparison with the modelling, chips A and B were fabricated from a Si-doped GaAs/AlGaAs heterostructure containing a 2DEG situated \SI{90}{nm} below the surface with density around \SI{1.5e11}{cm^{-2}} and mobility \SI{1.57e6}{cm^{2}V^{-1}s^{-1}}. Split gates were patterned by electron-beam lithography and metallised with Ti/Au (7/\SI{10}{nm}). A thick (\SI{1}{\mu\meter}) high-quality ZnO layer was deposited on chip B at room temperatures by a high-target-utilisation sputtering technique (HiTUS).\cite{garcia2010ultrafast, pedros2011guided} To avoid accompanying ion implantation, which greatly reduces the 2DEG conductance, a thin (\SI{20}{nm}) amorphous aluminium oxide buffer layer was deposited (by atomic-layer deposition) after the gate metallisation but before sputtering ZnO. At $T=\SI{4.2}{K}$, a source-drain current was driven by a \SI{0.1}{mV} bias and the conductance, $G$, was measured with a lock-in amplifier at \SI{77}{Hz}. Fig.~\ref{modelpinch}(a) and (b) demonstrate pinch-off characteristics of different 1D channels in chips A and B with different dielectric layers above the surface. At $V_{\rm SG}=\SI{-0.2}{V}$, the 2DEG under the gate is depleted, defining the 1D channel, 
\begin{figure*}
\centering
	\includegraphics[scale=1.15]{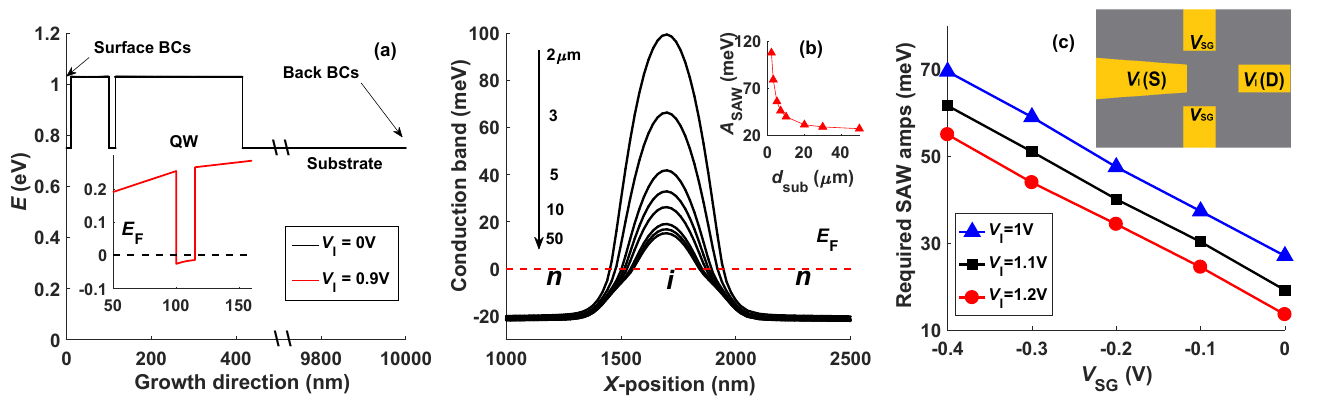}
	\caption{\label{fig:simundoped}(a) Conduction-band profile in an undoped GaAs/AlGaAs heterostructure with an inducing-gate voltage $V_{\rm I}$ of \SI{0}{V} and (Inset) \SI{0.9}{V}. At $V_{\rm I}=\SI{1}{V}$, a 2DEG is induced in the QW. (b) Potential energy profile through a lateral $n$-$i$-$n$ junction with different depths to the back surface $d_{\rm sub}=2,3,5,7,10,20,30$ and $\SI{50}{\mu\meter}$ at $V_{\rm I}=\SI{1.2}{V}$ and $V_{\rm SG}=0$. The Fermi level is taken to be 0. Inset: the required SAW amplitude $A_{\rm SAW}$ as a function of $d_{\rm sub}$. (c) Required SAW amplitude ($d_{\rm sub}=\SI{50}{\mu\meter}$) as a function of $V_{\rm SG}$ at different $V_{\rm I}$ of \SI{1}{V} (blue triangle), \SI{1.1}{V} (black square), and \SI{1.2}{V} (red circle). Inset: Schematic diagram of an induced $n$-$i$-$n$ junction.}
\end{figure*}
and the channel is finally pinched off at $V_{\rm SG}=V_{\rm P}$. $V_{\rm P}$ is dependent on the length of 1D channel, becoming more negative for a shorter lithographic channel length $L$ (due to fringing fields at the ends of the channel). In chip B, the ZnO overlayer leads to a $\sim\SI{0.25}{V}$ shift of $V_{\rm P}$ towards zero, compared with chip A. To exclude the possibility of `damage' to the 2DEG during the deposition, we etched away the ZnO layer on chip B with 20\% HCl solution (chip B' in Fig.~\ref{modelpinch}(b)). $V_{\rm P}$ became comparable to the values for chip A (the remaining slight difference can be explained by the presence of the thin Al$_2$O$_3$ buffer layer). As shown in Table~\ref{tab:table1}, the experimental results for $V_{\rm P}$ match the calculation very well, in particular having the same shift when there is a ZnO layer. This shows that it is essential to take into account the effect of the dielectric layer.

There are limitations in comparing theory and experiment. Experimentally, for a 1D channel, $V_{\rm P}$ is affected by wafer disorder, lithographic imperfections, device cool-down rate, sweep direction and sweep rate. There are also uncertainties in the dielectric constant and Schottky barrier energy for ZnO grown by HiTUS, as they are dependent on surface conditions, crystal quality, etc.\cite{brillson2011zno}. In MBE growth, there is a low but uncertain density of unintentional dopants in heterostructures, typically $p$-type from carbon atoms. In Table \ref{tab:table1} we compare $V_{\rm P}$ with and without \SI{e13}{cm^{-3}} fully-ionized $p$-type dopants, and the errors indicate the spread between these two cases. However, the error in $V_{\rm P}$ caused by this uncertainty is much less than the measured one with and without the ZnO overlayer, proving the importance of the boundary condition at the surface.

\section{Boundary conditions in undoped gated devices}
For a more sensitive test of BCs, we consider a second type of gated device on an undoped GaAs/AlGaAs heterostructure. Fig.~\ref{fig:simundoped}(a) illustrates the conduction band of such a heterostructure. An external electric field applied by a surface inducing gate pulls the conduction (valence) band below (above) the Fermi level, inducing free electrons (holes) inside a quantum well (QW). The carrier density of this 2D gas is tuned by gate bias instead of doping density. This inducing technique greatly reduces the density of ionised scatterers and gives a high carrier mobility with a low carrier density.\cite{harrell1999very} With a positive bias on two inducing gates ($V_{\rm I}>\SI{0.8}{V}$) separated by \SI{600}{nm}, electrons accumulate under each inducing gate, forming a source and drain separated by an intrinsic barrier (a lateral $n$-$i$-$n$ junction). The electrostatic potential distribution through the $n$-$i$-$n$ junction is calculated using our model and verified experimentally by using a surface acoustic wave (SAW) to pump electrons across the potential hill in the intrinsic region. In the classical SAW-pumping mechanism, if the maximum downward slope in the SAW potential is greater than that in the approach to the potential hill, electrons can be confined in SAW minima and dragged across the potential hill to the drain. If not, SAW minima flatten out before reaching the point with the maximum slope on the hill, and all electrons are pushed back to source.\cite{robinson2001classical,kataoka2006experimental} Given that a SAW is a sine wave, the minimum required SAW amplitude is estimated from the electrostatic potential through the $n$-$i$-$n$ junction. In a real device the applied amplitude has to be larger because of screening by gates. Unlike for a doped device, there is no built-in electric field in an undoped device, and so there are no frozen charge layers at either front or back interfaces. This makes the choice of BCs more critical in an undoped device than the doped one.

To induce a 2D gas, a voltage is applied to an inducing gate, causing a large electric field below the surface as well as outside the surface. Thus a calculation of the potential should include a dielectric layer and set a Neumann BC at the top of it, as described above. In the model of a doped device, we found that the position of the back BC is not important. However the lack of back-charge states in an undoped device can help us to probe the back BC. If one assumes that the bands are pinned at the regrowth interface $\sim\SI{2}{\mu\meter}$ below the surface, the maximum potential slope in the $n$-$i$-$n$ junction is so large that a SAW with amplitude greater than \SI{100}{meV} appears to be required to pump electrons. On a GaAs substrate for similar SAW devices, the SAW amplitude is measured around 20--\SI{30}{meV} at the power of 8--\SI{10}{dBm}.\cite{schneble2006quantum,naber2006surface,kataoka2006experimental} 
Given this, it is impossible to realise SAW pumping in a such an induced $n$-$i$-$n$ junction, which is in conflict with our experiment observation. Fig.~\ref{fig:simundoped}(b) illustrates the potential through the $n$-$i$-$n$ junction for various depths $d_{\rm sub}$ of the back BC. As $d_{\rm sub}$ increases, the required SAW amplitude decreases significantly and stabilises at around \SI{25}{meV} for depths over \SI{20}{\mu\meter} (Inset to Fig.~\ref{fig:simundoped}(b)). This saturation depth is strongly dependent on the dimension of intrinsic region. A larger $d_{\rm sub}$ is required for accurate calculation of a longer intrinsic channel in an $n$-$i$-$n$ junction.

To manipulate the SAW pumping process in our model, a pair of split gates is placed on the sides of the intrinsic region (Inset to Fig.~\ref{fig:simundoped}(c)). We calculate the electrostatic potential across the induced $n$-$i$-$n$ junction at different $V_{\rm SG}$ and $V_{\rm I}$, and from the maximum slope of the potential hill the required SAW amplitudes are estimated, as shown in Fig.~\ref{fig:simundoped}(c). As $V_{\rm SG}$ becomes more negative, the channel is squeezed and the intrinsic potential hill becomes higher, requiring larger SAW amplitude. When increasing $V_{\rm I}$, the induced 2DEG density and Fermi energy increase and the source and drain regions expand, pulling down the potential hill so that the slope above the Fermi level is less, requiring lower SAW amplitude. In the pumping process, SAWs provide the longitudinal confinement and bias on the split gates provides the transverse confinement. The two together define a dynamic quantum dot (DQD) in each SAW minimum, containing a precise number of electrons, $n$, as the Coulomb charging energy is sufficient to prevent confinement of an extra electron. This has been shown to give a quantised acoustoelectric current $I=nef$, where $f$ is the SAW frequency.\cite{1997single,ford2017transporting}

\begin{figure}
\centering
\includegraphics[scale=0.8]{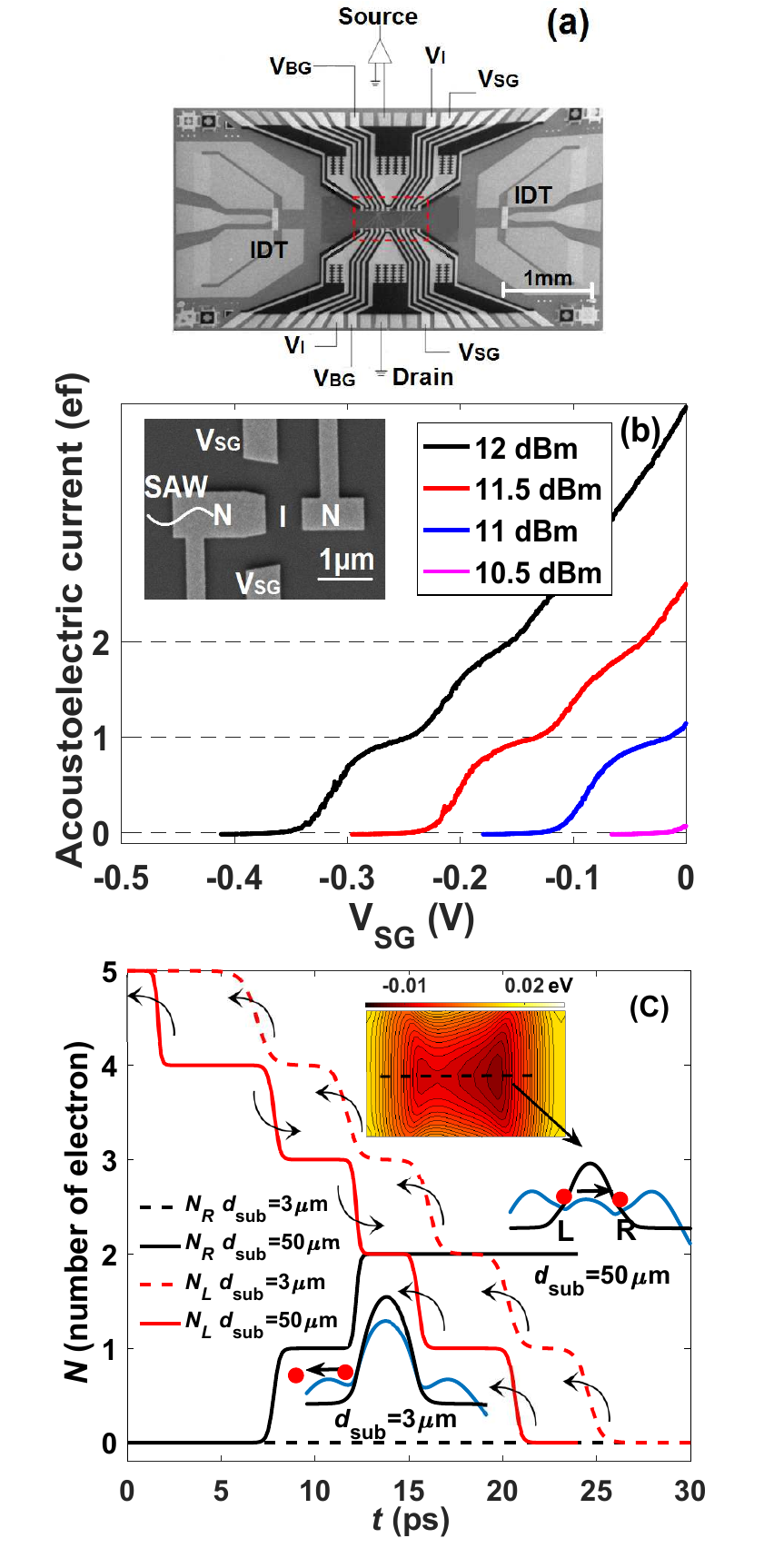}
\caption{ (a) Schematic diagram of the device design. (b) Quantised SAW-driven current $I_{\rm SAW}$ in a lateral $n$-$i$-$n$ junction ($V_{\rm I}=\SI{1.2}{V}$) as a function of $V_{\rm SG}$ with different RF powers. Inset: SEM image of the junction. (c) Calculations matching experimental device for $d_{\rm sub}=\SI{50}{\mu m}$ and \SI{3}{\mu m}: a SAW (amplitude \SI{25}{meV}) is superimposed on the potential ($V_{\rm I}=\SI{1.2}{V}$, $V_{\rm SG}=\SI{0}{V}$). Minima form on either side of the intrinsic barrier, and the curves show the numbers $N_{\rm L}$ and $N_{\rm R}$ of electrons in them as a function of time as the SAW moves to the right. Arrows after each step show the direction in which that electron tunnels out of the left dot. Inset: 2D combined potential (top) and 1D profiles along the junction with (blue) and without (black) the SAW.}
	\label{inducedresult}
\end{figure}
\section{SAW dynamic quantum dots}
Fig.~\ref{inducedresult}(a) illustrates the induced device  in an undoped GaAs/AlGaAs heterostructure with a lateral $n$-$i$-$n$ junction, matching the modelled device. On a thick GaAs substrate, a \SI{35}{nm} GaAs QW is sandwiched between two thick AlGaAs layers, with a thin GaAs capping layer on top. A \SI{1}{\mu m} wavelength SAW with $f=\SI{2.8}{GHz}$ is launched by an inter-digital transducer. At a low temperature of $T=\SI{4.2}{K}$ and inducing-gate voltage $V_{\rm I}=\SI{1.2}{V}$, a strong SAW overcomes the potential hill in the intrinsic region and drags electrons from source to drain, exhibiting quantised acoustoelectric current as a function of $V_{\rm SG}$ at different SAW powers from 10.5--\SI{12}{dBm}, as shown in Fig.~\ref{inducedresult}(b). This is the first time quantised SAW-driven current has been observed in an induced device. The dashed lines in Fig.~\ref{inducedresult}(b) show the expected positions of the first two plateaux at $I=ef\sim\SI{0.45}{nA}$ and $I=2ef\sim\SI{0.90}{nA}$. The threshold pinch-off voltage $V_{\rm P}$ ranges from \SI{-0.12}{V} at a SAW power of \SI{11}{dBm} to \SI{-0.33}{V} at \SI{12}{dBm}, which corresponds to a change in amplitude by a factor of 1.12. The model in Fig.~\ref{fig:simundoped}(c) shows the same trend in $V_{\rm P}$ but such a change in $V_{\rm P}$ requires a larger increase of SAW amplitude than in the experiments. This can probably be explained by a charging effect that caused a drift of $V_{\rm P}$ in this sample. We notice that the first plateau is visible at $V_{\rm SG}=\SI{0}{V}$ (at a power of \SI{11}{dBm}), which shows that there is strong transverse confinement even with a grounded split gate---the split gates screen the field from the inducing gates, so that, close to the side gates, the bands are not as near the Fermi energy. Our calculation at $V_{\rm SG}=\SI{0}{V}$ also gives transverse confinement, which fits a parabolic potential with energy-level spacing \SI{1}{meV}. This shows that grounded gates and exposed surfaces behave differently in undoped material, whereas in doped heterostructures there is usually very little difference.

In order to model the quantised acoustoelectric current, we take a simple model in which electrons in a SAW-driven dot are able to tunnel out via saddle-point potential barriers. The transmission probability through such a barrier with potential $V(x,y)=V_0-\frac{1}{2}m^*\omega_x^2+\frac{1}{2}m^*\omega_y^2$ is \cite{buttiker1990quantized,aizin1998screening}
$$ T=\frac{1}{1+e^{-\pi\epsilon}}$$
where $$\quad \epsilon=\frac{2(E_{N}-\frac{1}{2}\hbar\omega_y-V_0)}{\hbar\omega_x},$$
and $m^*$ is the effective mass. We superpose a SAW potential 
on to the calculated electrostatic potential and calculate $\omega_x$ for the barriers, together with the energy of the $N^{\rm th}$ electron, $E_{N}$, which is estimated from the electron ground-state energy in the SAW minimum and a constant Coulomb charging energy, taken to be \SI{3}{meV}.\cite{astley2007energy}.

In studies of quantised SAW pumping in doped devices, only back tunnelling is usually considered.\cite{ford2017transporting} However, for the short intrinsic channel used in this experiment, we need to calculate tunnelling processes through both the back and front barriers. Our model shows that it is still possible, and likely, that electrons in the SAW minimum, which we label L, will tunnel forwards into the minimum ahead of the intrinsic barrier, which we label R, provided that we use deep BCs. 
Fig.~\ref{inducedresult}(c) shows the numbers of electrons $N_{\rm L}$ and $N_{\rm R}$ in the minima, for two different back boundary conditions.

Firstly, for a deep BC, $d_{\rm sub}=\SI{50}{\mu m}$ (solid lines), during a SAW cycle,  $N_{\rm L}$ decreases as electrons tunnel back to the source through the back barrier. However, at some point (around \SI{7}{ps} in the plot) the probability of tunnelling forwards through the front barrier becomes greater than that of going backwards, causing the confinement to decrease and the electrons trapped in the dot to tunnel forwards, increasing $N_{\rm R}$ (upper inset). Later, at around \SI{15}{ps} here, the right dot starts rising up again so that forward tunnelling stops, and back tunnelling starts again. 
Eventually the left dot empties. This results in an integer number of electrons being pumped through the intrinsic region in each SAW cycle, yielding a quantised current.

In contrast, for shallow BCs ($d_{\rm sub}=\SI{3}{\mu m}$, shown with dashed lines), the front barrier is so high that all electrons in the SAW minima tunnel back to the source through the back barrier (lower inset). Therefore $N_{\rm R}=0$ over a whole SAW cycle, which does not match the experiment. In reality, metal gates and free charges screen and attenuate the SAW,\cite{aizin1998screening} whereas we assume a constant SAW amplitude in the above model. Screening would only reduce the chance of pumping for a given applied SAW amplitude, so with shallow back BCs it would still be impossible to pump electrons. Deep BCs are vital to explain our experimental observation of pumping, and this highlights the important role of freezing of charge at the regrowth interface in patterned devices on undoped (and doped) heterostructures, which has largely been ignored in the past.
\section{Conclusion}
To conclude, we have compared experiments on gate-patterned quantum devices at cryogenic temperatures with self-consistent electrostatic modelling using various boundary conditions and standard software. The models are fairly accurate, provided that the boundary conditions are chosen carefully. For real 1D channels in a doped GaAs heterostructure, the pinch-off voltage 
shifts 
significantly as the dielectric layer on the surface is changed from vacuum to ZnO. In order to account for this in our model, we have to treat the front surface of the GaAs as having a frozen charge layer, rather than simply making it satisfy particular boundary conditions below the dielectric layer. To refine the back BCs, we compared modelling and experiments on pumping electrons through an induced lateral $n$-$i$-$n$ junction in an undoped GaAs heterostructure using a surface acoustic wave. We find that it is important to move the back boundary much deeper than the MBE regrowth interface, which has often been taken as an equipotential. With these improved boundary conditions, it is possible to accurately model and optimise complex gate-patterned quantum devices.

We acknowledge the support of the Cambridge International Trust (HH and TKH) and the China Scholarship Council (HH). 

\bibliography{iopart-num2}
\end{document}